\documentclass[11pt]{JHEP3}
\usepackage{epsfig}
\usepackage{amsmath}

\newcommand{\fnl}{f_\mathrm{NL}}
\newcommand{\gnl}{g_\mathrm{NL}}
\newcommand{\zetag}{\zeta_\mathrm{g}}
\newcommand{\fsky}{f_\mathrm{sky}}

\title{Estimating \boldmath $\fnl$ and $\gnl$ from Massive High-Redshift Galaxy Clusters}

\author{Kari Enqvist\footnote{E-mail: kari.enqvist@helsinki.fi}, Shaun Hotchkiss\footnote{E-mail: shaun.hotchkiss@helsinki.fi}, Olli Taanila\footnote{E-mail: olli.taanila@iki.fi}\\Department of Physics, University of Helsinki\\Helsinki Institute of Physics}

\abstract{There are observations of at least 15 high-redshift massive galaxy clusters, which
have an extremely small probability with a purely Gaussian initial curvature perturbation. Here we revisit the estimation of the contribution of non-Gaussianities to the cluster mass function and point out serious problems that have resulted from the application of the mass function out of the range of its validity. We remedy the situation and show that the values of $\fnl$ previously claimed to completely reconcile (i.e. at $\sim100\%$ confidence) the existence of the clusters with $\Lambda$CDM are unphysically small. However, for WMAP cosmology and at 95\% confidence, we arrive at the limit $\fnl \gtrsim 411$, which is similar to previous estimates. We also explore the possibility of a large $\gnl$ as the reason for the observed excess of the massive galaxy clusters. This scenario, $\gnl > 2\times10^6$, appears to be in more agreement with CMB and LSS limits for the non-Gaussianity parameters and could also provide an explanation for the overabundance of large voids in the early universe.}

\preprint{HIP-2010-35}


\begin{document}

\section*{Introduction}

Many models of inflation predict an unobservable non-Gaussianity of the primordial curvature perturbation $\zeta$. Thus a detection of primordial non-Gaussianity at any level would rule out whole classes of models. Although observationally the primordial perturbations are Gaussian to a great accuracy, there is nevertheless still much room for non-Gaussianities,
which are usually parameterized by the lowest-order non-linearity parameters $\fnl$ and $\gnl$. They can be defined by an expansion around a Gaussian perturbation $\zeta_\mathrm{g}$ with
\begin{equation}\label{eq:fnlgnldef}
\zeta = \zetag + \frac{3}{5}\fnl(\zeta_\mathrm{g}^2-\langle \zeta_\mathrm{g}^2 \rangle)+\frac{9}{25}\gnl\zeta_\mathrm{g}^3+\mathcal{O}(\zeta_\mathrm{g}^4) \; .
\end{equation}
This expansion assumes a specific local form of non-Gaussianity. The non-linearity parameters have previously been constrained by both CMB and LSS measurements.  The best
constraints for $\fnl$ come from WMAP 7-year results \cite{WMAP7} $-10 < \fnl^\mathrm{local} < 74$. For comparision, many inflaton models predict $|\fnl|\lesssim {\mathcal O}(1)$. For $\gnl$ the limits are less strict, with WMAP 5-year results giving \cite{VielvaSanz}
$-5.6\times10^5<\gnl<6.4\times10^5$, while halo bias and LSS yield \cite{DesjacquesSeljak2009} $-3,5\times10^5<\gnl<8.2\times10^5$. For the latter limits one assumes that $\fnl \sim 0$. It is noteworthy that the estimates from CMB and LSS for $\gnl$ are comparable. The expectation is that the Planck Surveyor Mission should be able to limit $\fnl\lesssim {\mathcal O}(5)$ \cite{Planck}.

Recently, considerations of primordial non-Gaussianity have been extended
to studies of high mass galaxy clusters. The interest has been triggered by
the fact that at least 15 high-redshift ($z>1.0$) galaxy clusters have been
observed with masses measured to be about
$\sim10^{14}\mathrm{M}_\odot$ \cite{toobigtooearly}. These clusters
have redshifts in the range $1.02 \leq z \leq 1.62$ while the central
values of their masses lie the range
$0.57\times10^{14}\mathrm{M}_\odot \leq m \leq
1.0\times10^{15}\mathrm{M}_\odot$
\cite{data1,data2,data3,data4,data5,data6,data7,data8}. It has been argued that in the standard
$\Lambda\mathrm{CDM}$-cosmology with purely Gaussian initial
perturbations, the probability for the existence of such very high-mass clusters
is diminishingly small \cite{Jeeetal}\footnote{It is also argued in \cite{Mortonson:2010mj} that a more conservative treatment of survey volume and measurement bias could reconcile these clusters with $\Lambda$CDM; however this conclusion is only arrived at for the clusters individually and not the ensemble as a whole.}.

The situation changes in the presence of primordial non-Gaussianity.
Since $\fnl$ and $\gnl$ are the coefficients of products of several
Gaussian variables, they modify the behaviour  of the PDF in the
tails of the distribution, non-zero $\fnl$ increasing the
probability of massive clusters and decreasing the probability for
large voids, and $\gnl$ increasing the probability of both. Thus if
interpreted as a primordial feature, the anomalous abundance of
large massive clusters might be evidence for departure from the
Gaussianity of the primordial perturbation \cite{toobigtooearly,CayonGordonSilk}.

The effect of non-Gaussianity on the abundance of massive clusters is encoded in the cluster mass function $n(M,z,\fnl,\gnl)$. However, deriving $n$ is a very involved calculation already in the Gaussian case. When non-Gaussianities are present, certain problems arise in the use of the cluster mass function, which, if not accounted for properly, result in large errors in the final estimates for the non-linearity parameters. Here we present a careful examination of the influence of non-Gaussianities on the cluster mass function, resolving the discrepancies that exist in the literature, and finding what we believe is the correct observational lower limit on $\fnl$ from the observed number of massive clusters. Moreover, we extend the previous analyses to higher order statistics and find the cluster limit also on $\gnl$. We argue that a large $\gnl$ is in more agreement with the CMB and LSS limits and point out that it could also provide an explanation for the observed overabundance of large voids in the early universe (see \cite{Voids1,Voids2,D'Amico:2010kh} and references therein).

The paper is organized as follows. In section \ref{section:method} we introduce the formalism necessary to perform the quantitative analysis.  In section \ref{section:analytical} we then derive analytical estimates for the skewness and the kurtosis of the spectrum of density perturbations for given $\fnl$ and $\gnl$. After that, in section \ref{section:fnl} we estimate the value of $\fnl$ required to explain the number of heavy clusters using both analytical estimates and numerical calculations. We also discuss the subtleties leading to unreliable underestimates of $\fnl$ found in the literature. In section \ref{section:gnl} we then repeat this analysis assuming that the contribution of $\fnl$ is insignificant, and that the non-Gaussian primordial statistics is dominated by $\gnl$. In section \ref{section:conclude} we summarize our results and point out the virtues of large $\gnl$ in providing a possible explanation for the apparent overabundance of voids in the universe.

\TABLE[t]{
\begin{tabular}{r c c c}
\textbf{Cluster Name} & \textbf{Redshift} & \boldmath $M_{200}\,10^{14}M_\odot$ & \textbf{Mass Reference}\\
\hline
WARPSJ1415.1+3612 & $1.02$ & $3.33\substack{+2.83 \\ -1.80}$ & \cite{data3}\\
SPT-CLJ2341-5119 &$1.03$ & $5.40\substack{+2.80\\-2.80}$ & \cite{data2}\\
CLJ1415.1+3612 & $1.03$ & $3.40\substack{+0.60\\-0.50}$ & \cite{data6} \\
XLSSJ022403.9-041328 & $1.05$ & $1.66\substack{+1.15\\-0.38}$ & \cite{data4} \\
SPT-CLJ0546-5345 & $1.06$ & $10.0\substack{+6.00\\-4.00}$ & \cite{data1} \\
SPT-CLJ2342-5411 & $1.08$ & $2.90\substack{+1.80\\-1.80}$ & \cite{data2} \\
RDCSJ0910+5422 & $1.10$ & $6.28\substack{+3.70\\-3.70}$ & \cite{data5} \\
RXJ1053.7+5735(West) & $1.14$ & $2.00\substack{+1.00\\-0.70}$ & \cite{data7} \\
XLSSJ022303.0043622 & $1.22$ & $1.10\substack{+0.60\\-0.40}$ & \cite{data7} \\
RDCSJ1252.92927 & $1.23$ & $2.00\substack{+0.50\\-0.50}$ & \cite{data5} \\
RXJ0849+4452 & $1.26$ & $3.70\substack{+1.90\\-1.90}$ & \cite{data5} \\
RXJ0848+4453 & $1.27$ & $1.80\substack{+1.20\\-1.20}$ & \cite{data5} \\
XMMUJ2235.3+2557 & $1.39$ & $7.70\substack{+4.40\\-3.10}$ & \cite{data7} \\
XMMXCSJ2215.9-1738 & $1.46$ & $4.10\substack{+3.40\\-1.70}$ & \cite{data7} \\
SXDF-XCLJ0218-0510 & $1.62$ & $0.57\substack{+0.14\\-0.14}$ & \cite{data8} \\
\hline
\end{tabular}
\caption{A list of the 15 observed high-mass and high-redshift galaxy clusters that we use in the present analysis. This list was first compiled in \cite{toobigtooearly}.}
\label{ourtable}
}

\section{Method}
\label{section:method}

The theoretical Gaussian mass function was first calculated by
using a spherical collapse model \cite{PressSchechter}, and was
later improved by generalizing this to ellipsoidal collapse. Even
so, the theoretical predictions do not match the results of
simulations, and the best estimates for $n_\mathrm{G}(M,z)$ still
come from semi-analytical fits to N-body simulations.

The case for the non-Gaussian cluster mass function is even more complex. The N-body simulations have been mostly performed for Gaussian distributions, and thus no N-body formula for general non-Gaussian distribution exists. If deriving the accurate mass function was difficult for the Gaussian case, the non-Gaussianity of the distribution makes the task even more cumbersome. Thus it is no surprise that even the best estimates for the non-Gaussian mass function are only roughly valid, and do not always match the N-body simulations even in the Gaussian limit ($\fnl\to0$, $\gnl\to0$).

Since the analytical derivations have difficulty including non-ellipsoidal collapse (however, see \cite{MaggioreRiotto1,MaggioreRiotto2,MaggioreRiotto3,Simone}), and thus the non-Gaussian mass functions cannot be trusted directly, the ratio of the non-Gaussian to Gaussian mass functions $\mathcal{R}$ is often used to estimate the effect of non-Gaussianity,
\begin{equation}
\label{eq:ratio}
\mathcal{R}\left( M,z,\fnl,\gnl \right) = \frac{ n_\mathrm{analytical}(M,z,\fnl,\gnl) }{ n_\mathrm{analytical}(M,z,\fnl=0,\gnl=0)} \; ,
\end{equation}
so that the non-Gaussian mass function is given by the Gaussian mass
function (e.g.~a fit to the Gaussian  N-body simulation) multiplied
by the ratio,
$n_\mathrm{NG}(M,z,\fnl,\gnl)=n_\mathrm{G}(M,z)\,\mathcal{R}$. This
expression has the correct behaviour in the non-Gaussian limit, and
the erroneous behaviour of the non-Gaussian estimates is furthermore
assumed to cancel in the ratio. There is evidence from N-body
simulations that although apparently ad-hoc, these assumptions do
have some validity
\cite{Wagner:2010me,Grossi:2009an,DesjacquesSeljak2009}.

To estimate the value of the non-Gaussianity parameters, two functions are needed for the use of formula
(\ref{eq:ratio}): the numerical Gaussian mass function and the analytical non-Gaussian mass function.
For the non-Gaussian function there are several different expressions in the literature,
with the most common being the Matarrese-Verde-Jimenez estimate \cite{MVJ} (hereafter called the MVJ estimate). Other expansions include the Edgeworth
expansion \cite{edgeworth,ChongchitnanSilk} and its generalisations \cite{MaggioreRiotto1,MaggioreRiotto2,MaggioreRiotto3}.
Analysis of the applicability of the various expansions can be found in ref.~\cite{D'AmicoMussoNorenaParanjape}
who combine MVJ with the methods of \cite{MaggioreRiotto1,MaggioreRiotto2,MaggioreRiotto3}.
For the ranges of redshifts and masses that we need to consider here, however, the Edgeworth
expansion no longer converges. For that reason, we also use the MVJ expressions for both the
dependence on $\fnl$ and $\gnl$ \cite{DesjacquesSeljak2009}.

We will follow the MVJ \cite{MVJ} convention for defining the ratio $\mathcal{R}(M,z,\fnl,\gnl)$ by
\begin{equation}\label{eq:MVJfnl}
 \mathcal{R}
 =\exp\left(\delta_{ec}^3\frac{S_3(\fnl)}{6\sigma_M^2}+\delta_{ec}^4\frac{S_4(\gnl)}{24 \sigma^2_M}\right) \left\{ \frac{1}{6} \frac{\delta_{ec}}{\delta_3} \frac{dS_3}{d\ln \sigma} +\delta_3 \right\}\left\{\frac{1}{24}\frac{\delta_{ec}^2}{\delta_4} \frac{dS_4}{d\ln\sigma}+\delta_4 \right\}
\end{equation}
where $\delta_{ec}=\sqrt{0.75}\times 1.686$ is the critical density of ellipsoidal collapse, $\delta_3=\sqrt{1-\delta_{ec}S_3/3}$ and
$\delta_4=\sqrt{1-\delta_{ec}^2 S_4/12}$. $S_3(M,\fnl)$, $S_4(M,\gnl)$ and $\sigma_M$ all defined in section \ref{section:analytical}, are the smoothed, skewness, kurtosis and variance of the nearly Gaussian density perturbations, respectively.

The other popular definition for $\mathcal{R}$ uses the Edgeworth expansion, but is not suitable for our purposes for reasons described in \cite{D'AmicoMussoNorenaParanjape}. Specifically, the Edgeworth expansion involves an expansion over the terms appearing in the exponentials above. Despite the inherent smallness of $S_3$ and $S_4$, the terms involving $\nu=\delta_{ec}/\sigma_M$ can (and do) become large enough to overcome this. The Edgeworth expansion convention has been well tested in N-body simulations \cite{Wagner:2010me,Grossi:2009an} and performs well; however we will need to use our mass function beyond the ranges of these simulations. In \cite{MaggioreRiotto3} a full theoretical, non-Gaussian mass function is accurately determined that returns the
Edgeworth expansion for $\mathcal{R}(\fnl)$; however it is explicitly assumed that $\nu^3 \sigma_M S_3 \ll 1$ to arrive at this result, making the result correct, but not applicable to the very high mass clusters. In \cite{D'AmicoMussoNorenaParanjape} this condition is relaxed and the full non-Gaussian mass function is calculated taking this quantity into account non-perturbatively. The $\mathcal{R}$ that results from this mass function is equivalent to eq.~(\ref{eq:MVJfnl}) up to the accuracy with which we know the masses of the clusters studied in this work. It is expected that this mass function will break down when $\nu \sigma_M S_3 \simeq 1$, a condition we need to be aware of when integrating to high masses, but that does not hold for any of the masses and redshifts of the clusters themselves for $\fnl\lesssim 1000$. For $\nu \sigma_M S_3 > 1$ (i.e. $\fnl>1000$ and $\nu>5$) there are currently no N-body simulations or theoretical results available. To derive \emph{upper} bounds on the value of $\fnl$ allowed by clusters this problem will need to be addressed in the future.

From $n(M,z,\fnl,\gnl)$ we then calculate the expected number of
clusters of a given mass range, over a  given redshift interval and
with a given sky coverage using the following formula:
\begin{equation}\label{eq:defExp}
\mathrm{Exp}(M_\mathrm{range},z_\mathrm{range},\fnl,\gnl)=\int^{z_f}_{z_n}dz \int^{M_{\mathrm{max}}}_{M_{\mathrm{min}}} dM\, \fsky \frac{dV}{dz} n_\mathrm{NG},
\end{equation}
where $\fsky$ is the fractional sky coverage and $dV(z)/dz$ is the volume element at redshift $z$.

For both $\fnl$ and $\gnl$ we quote results using the CMB
convention. We also use WMAP5 \cite{Hinshaw:2008kr} cosmological
parameters\footnote{We do not use WMAP7 cosmology because the N-body
simulations to date have only been performed using WMAP5 parameters.
The deviations in our results arising from this will be much less
than the errors implicit in the method itself.}:
$h=0.705$, $n_s=0.96$, $\Omega_m=0.28$, $\Omega_b h^2=0.0227$ and
$\sigma_8=0.812$.

\section{Analytical estimate for \boldmath $S_3$ and $S_4$}
\label{section:analytical}

Due to a variety of differing claims in the literature, before
performing the numerical analysis to extract  information relating
to $\fnl$ and $\gnl$, it will be useful to have some approximate,
analytical, estimates for the effects both these parameters will
have. We start by estimating the sizes of $\sigma S_3$ and $\sigma^2
S_4$. These quantities are central to our calculation and different values have been claimed for them in the
literature using identical definitions.

\subsection[Estimation of $\alpha_R$ and $\sigma_R$]{Estimation of \boldmath$\alpha_R$ and $\sigma_R$}

Before we can estimate $S_3$ and $S_4$ we will first estimate the variance of the density perturbations smoothed over a radius $R$. This will be useful for testing our approximations and will make the eventual estimations of $S_3$ and $S_4$ much more straightforward. The variance, $\sigma_R$, is defined by 
\begin{equation}\label{eq:sigdef}
 \sigma_R^2=\int_0^{\infty} \frac{dk}{k} \alpha_R^2 (k,z) \mathcal{P}(k),
\end{equation}
where $\mathcal{P}$ is the almost scale invariant primordial power spectrum of $\zeta$, and 
\begin{equation}\label{eq:defalpha}
 \alpha_R (k,z) = \frac{2}{5 \Omega_m} D(z) \left ( \frac{k}{H_0} \right )^2
T(k) W_R (k).
\end{equation}
$H_0$ is the Hubble rate now, $D(z)$ is the linear growth function, $T(k)$ is the transfer function. For all results in this work we use the transfer function of \cite{Bardeen:1985tr} with the modified shape parameter of \cite{Sugiyama:1994ed}. The window function is given by
\begin{equation}
 W_R (k) = 3 \left(\frac{\sin(kR)}{(kR)^3 }- \frac{\cos(kR)}{(kR)^2}\right),
\end{equation}
which is the Fourier transformation of a window function that is a top-hat
in real space. The window function $W_R (k)$  is peaked at $k=0$;
however the combination $k^2 W_R (k)$ in eq.~(\ref{eq:defalpha}) is
closely peaked around $kR\simeq 1$ with an amplitude $\propto
1/R^2$.  $T(k)$ is approximately one for the largest scales,
but decreases for smaller scales that re-entered the horizon during
the radiation dominated era. It does not decrease fast enough on
these scales to stop the peak of the whole integral from still
occurring at $kR\simeq1$.

It is customary to define the normalisation of the power spectrum $\mathcal{P}(k)$ through the
parameter $\sigma_8$, which is just $\sigma_R$ evaluated at $R=8
h^{-1}\mathrm{Mpc}$ and $z=0$. Given this and the above, it is possible to
write, to a
reasonable approximation, the following expression for
$\sigma_R$,
\begin{eqnarray}\label{eq:sigalph}
 \sigma_R &=& \alpha_R \sqrt{\mathcal{P}(1/R)}D(z) \\
     \label{eq:sigest} &=&\sigma_8 \left(\frac{8}{R}\right)^{2} \frac{T(1/R)}{T(1/8)} D(z),
\end{eqnarray}
where $\alpha_R$ is defined as $\alpha_R(1/R,0)$. Ostensibly
eq.~(\ref{eq:sigalph}) defines $\sigma_R$  as a function of
$\alpha_R$ and $\mathcal{P}$; however, when estimating $S_3$ and
$S_4$, we will find it more useful to use this equation as a means
to replace $\alpha_R$ by the two quantities $\sigma_R$ and
$\mathcal{P}$ that we know the approximate sizes of.

In figure \ref{fig:sigest} we plot eq.~(\ref{eq:sigest}) as well as a numerically calculated curve for a  range of $R$ values at $z=0$. Both curves will change identically with redshift through the linear growth function $D(z)$.

\EPSFIGURE[t]{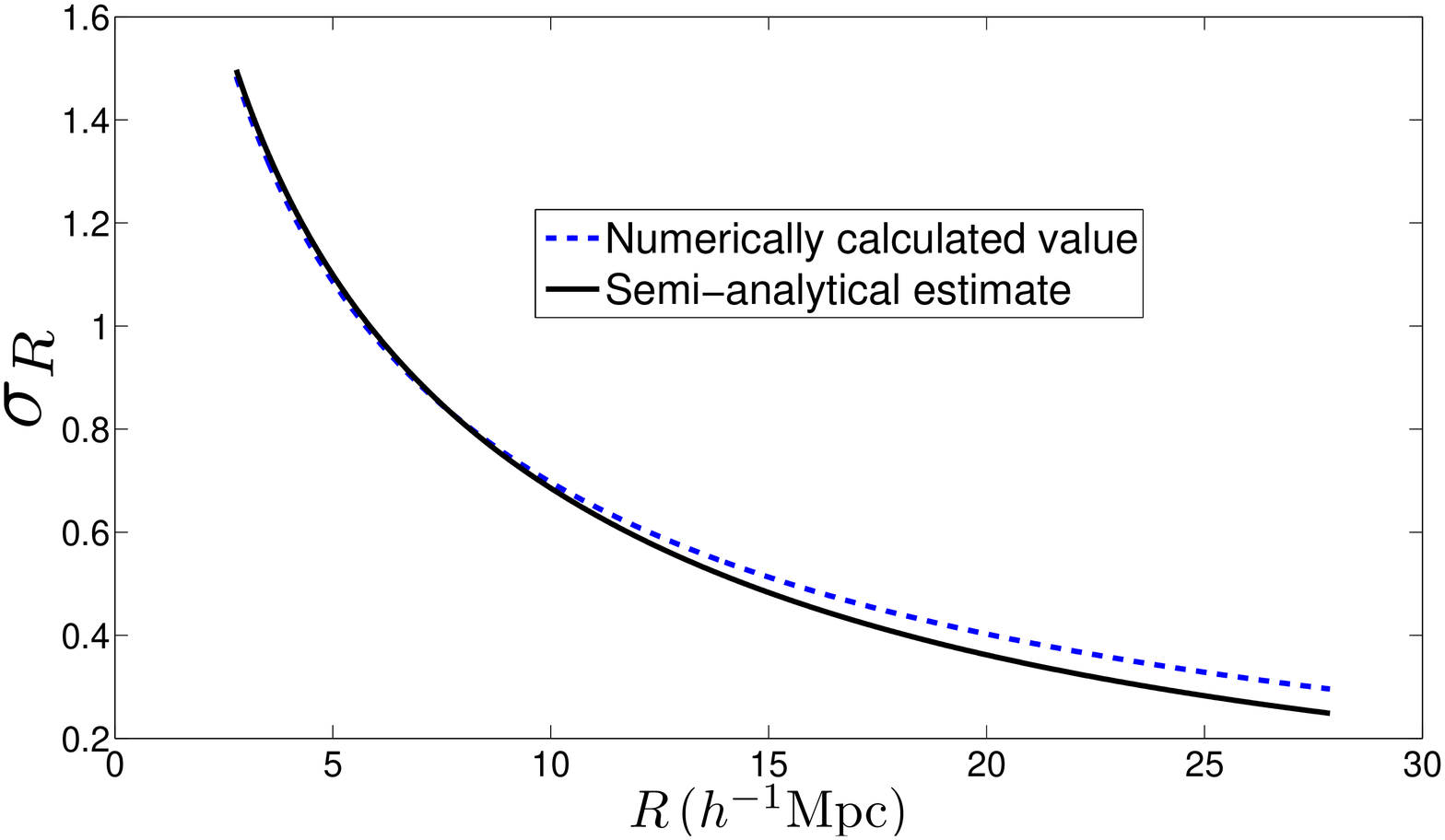,width=12cm}{\label{fig:sigest}$\sigma_R$ (at $z=0$) calculated numerically and using eq.~(\ref{eq:sigest}). The range of $R$ corresponds to a mass range of $M=10^{13}M_\odot - 10^{14}M_\odot$.}

\subsection[Estimation of $\sigma S_3$ and $\sigma^2 S_4$]{Estimation of \boldmath$\sigma S_3$ and $\sigma^2 S_4$}

Using the same arguments as above it is possible to estimate expressions for
$S_3$ and $S_4$. We
start with the skewness which is given by the following expression, using
$\fnl$ defined in eq.~(\ref{eq:fnlgnldef}),
\begin{equation}
 \sigma_R^4 S_3(R)=3 \fnl  \int_0^\infty \frac{dk_1}{k_1} \alpha_R(k_1)
\mathcal{P}_1 \int_0^\infty \frac{dk_2}{k_2} \alpha_R(k_2) \mathcal{P}_2\int^1_{-1} d\mu \, \alpha_R(k_3)\;,
\end{equation}
where $k_3^2 = k_1^2+k_2^2+2\mu k_1k_2$ and $\mathcal{P}_i$ is short for $\mathcal{P}(k_i)$.

The three factors of $\alpha_R(k)$ above will scale in the same way as they do
for $\sigma_R$. That is they will be highly peaked around the value $k=1/R$, with
magnitudes proportional to $1/R^2$. The integrals over $k_1$ and $k_2$ can be
evaluated in exactly the same manner as the $k$ integral for $\sigma_R$. The $\mu$
integral can be evaluated by appealing to the fact that $k_1$ and $k_2$ will be
$\simeq 1/R$ but not always exactly equal to $1/R$. Therefore, over the full range
$\mu=-1$ to $1$, it will always be possible to find values of $k_1\simeq 1/R$
and $k_2\simeq 1/R$ such that $k_3\simeq 1/R$. Therefore we can proceed by
also substituting $\alpha_R(k_3)$ with $\alpha_R$ and multiplying the rest of the
expression by 2 (the range of the $\mu$ integral). This gives,
\begin{equation}
\nonumber \sigma_R^4 S_3(R) = 6 \fnl \alpha_R^3 \mathcal{P}^{2}=
         6 \fnl \sigma_R^3 \mathcal{P}^{1/2}.
\end{equation}
Therefore, to the degree of our approximation, the quantity
$\sigma_R S_3(R)$ is scale independent and given by a very simple formula. For
$\sigma_8=0.812$, $\mathcal{P} \simeq 2.4\times 10^{-9}$ which allows us to make
the following estimate,
\begin{equation}\label{eq:S3est}
 \sigma_R S_3(R) \simeq 3 \times 10^{-4} \fnl.
\end{equation}
The growth factor $D(z)$ is independent of scale, therefore the combination $\sigma_R S_3(R)$ is exactly independent of redshift.

A very similar argument applies for the kurtosis, $S_4(R)$. The kurtosis is given by,\footnote{This particular expression for the kurtosis was first written down in an earlier version of ref.\cite{ChongchitnanSilk}.}
\begin{eqnarray}
 \nonumber \sigma_R^6 S_4(R) = \frac{3}{\pi} \gnl
\left(\prod^3_{i=1}\int^\infty_0 \frac{dk_i}{k_i} \alpha_R(k_i)
\mathcal{P}(k_i)\right) \\
\times \int^1_{-1} d\mu_1 \int^1_{-1} d\mu_2 \int^{2\pi}_0
d\phi\, \alpha_R(k_4),
\end{eqnarray}
with,
\begin{eqnarray}
\nonumber k_4^2=k_1^2+k_2^2+k_3^2+2k_1k_2\mu_1+2k_2k_3\mu_2 \\
\nonumber +2k_1k_3
\left(\cos{\phi}\sqrt{1-\mu_1^2}+\mu_1\mu_2\right).
\end{eqnarray}
If we follow the same process as we did for the skewness, this reduces to,
\begin{equation}
\sigma^6_R S_4(R) = 24 \gnl \alpha_R^4 \mathcal{P}^3 = 24 \gnl \sigma_R^4 \mathcal{P}.
\end{equation}
After we substitute the same value for $\mathcal{P}$ this gives,
\begin{equation}
\sigma_R^2 S_4(R) = 5.8\times 10^{-8} \gnl.
\end{equation}
As with $\sigma_R S_3(R)$, this is also exactly independent of redshift.

It was stated in an earlier version of this paper that our values of $\sigma_R S_3(R)$ and $\sigma_R^2 S_4(R)$ were similar to those given in refs.~\cite{DesjacquesSeljak2009,Desjacques:2008vf}, but very different to the ones given in ref.~\cite{ChongchitnanSilk}. The authors of ref.~\cite{ChongchitnanSilk} have revised their paper and all sets of results are now in agreement. The value of $S_3$ used in \cite{toobigtooearly} also matches our numerical result and analytic estimate.\footnote{Ben Hoyle, private correspondence} In our later numerical calculations we calculate $S_3(R)$ fully numerically. We do not do the same for $S_4(R)$. The implications of this are discussed briefly in Section \ref{sec:numgnl}. Our numerical calculation of $S_3$ matches our analytic estimate to within a factor $\sim 1-1.5$.

\section{Estimating \boldmath$\fnl$}
\label{section:fnl}

It was asked in \cite{toobigtooearly,CayonGordonSilk} ``what is the probability that a given cluster is the `most massive' cluster in the survey window?''. Here, the phrase `most massive' strictly speaking means \emph{least probable} because the redshift dependence of the mass function is always included. That is, a less massive cluster that collapsed much earlier could still be the `most massive' cluster by this definition. To answer this question both references create three mass bins. One mass bin contains the central mass of the cluster as well as all the masses contained within the 1-$\sigma$ error range either side of it. The second mass bin contains all the masses above the upper bound of this previous range. The probabilities are obtained by first calculating the expected number of clusters observed in each bin, then Poisson sampling from each of these distributions a large number of times $(10^4)$ and asking which bin contains the largest cluster each time. It is assumed that the third bin, containing masses beneath the 1-$\sigma$ error range always contains at least one cluster in any given survey. 

Both refs.~\cite{toobigtooearly,CayonGordonSilk} saw a sudden change in their calculated probabilities at a particular value of $\fnl$. This point occurs at different values of $\fnl$ in each reference. The main symptom of this change is a sudden leap in the probability that the upper mass range contains at least one observed cluster. The references interpret this as being caused by the mass around the lower limit of this range becoming more probable at this value of $\fnl$. It is striking that this transition occurs so rapidly in each reference and at considerably different values of $\fnl$. We believe the true cause of this effect is the breakdown of the mass function being used in each reference at a much larger mass, corresponding to the arbitrary upper bound in their numerical calculation of the integral in eq.~(\ref{eq:defExp}).

We numerically examine this effect in section \ref{sec:Gaussmass}; however in section \ref{sec:fnlest} below, following the spirit of section \ref{section:analytical}, we attempt to give an approximate analytic estimate of how large $\fnl$ would need to be to observe the magnitude of effect seen in refs.~\cite{toobigtooearly,CayonGordonSilk}.

\subsection[Estimate of $\fnl$ required to see the effect in ref.~\cite{toobigtooearly}]{Estimate of \boldmath$\fnl$ required to see the effect in ref.~\cite{toobigtooearly}}\label{sec:fnlest}
The cluster \textbf{XMMUJ2235.3+2557}, with mass $7.7^{+4.4}_{-3.1}\times 10^{14} M_\odot$, observed at redshift
$z=1.39$ (see table \ref{ourtable}) is found by \cite{toobigtooearly} and us to be one of the two least probable clusters. This cluster was detected by an X-ray survey \cite{Mullis:2005hp}. The full survey footprint of various X-ray surveys is estimated in \cite{toobigtooearly} to be 283 sq.~degrees. If we use the Jenkins et al mass function of ref.\cite{Jenkins} in eq.~(\ref{eq:defExp}), and then integrate from redshift $z=1.39 \rightarrow 2.2$ and from mass $M=12\times 10^{14} M_\odot \rightarrow \infty$ we find the expected number of clusters in this upper bin should be $0.0019$. This translates into a probability for this mass bin of $\simeq 0.002$ which is consistent with figure 2 of ref.~\cite{toobigtooearly}.

$\fnl$ will change this probability through the ratio in eq.~(\ref{eq:MVJfnl}) which to a good
approximation is,
\begin{equation}\label{eq:numvj}
 \mathcal{R}=\exp\left(\frac{\nu^3}{6} (\sigma S_3)\right)
\end{equation}
where we have made the useful substitution, $\nu=\delta_{ec}/\sigma_M$ to simplify the
result. To the degree of approximation we are going to make, all the scale and redshift dependence is
found in $\nu$ and $\sigma S_3$ is given by eq.~(\ref{eq:S3est}). 

We need to estimate the value of $\nu$ at $M=12\times 10^{14}M_\odot$ and redshift $z=1.39$. To do this, we will use eq.~(\ref{eq:sigest}) for $\sigma_M$ (note that $\delta_{ec}\simeq1.46$). Eq.	~(\ref{eq:sigest}) gives $\sigma_R$, therefore we need to find what smoothing scale $R(M)$ corresponds to each mass. We are interested in the comoving smoothing scale, therefore $R(M)$ is given by
\begin{equation}
 R=\left(\frac{3M}{4\pi\rho_m}\right)^{1/3},
\end{equation}
where $\rho_m$ is the \emph{present} density of matter in the Universe. For a critical density of $\rho_c=2.775/h \times10^{11}M_\odot \, (h\mathrm{Mpc}^{-1})^3$, $h=0.7$ and $\Omega_m=0.28$,
\begin{equation}
 R\simeq6M_{14}^{1/3} \, (h^{-1}\mathrm{Mpc}),
\end{equation}
where $M_{14}$ is the mass of the cluster in fractions of $10^{14} M_\odot$. Therefore, for $M_{14}=12$, we arrive at $R\simeq 13.7$. Upon direct substitution into eq.~(\ref{eq:sigest}) this gives for $\nu$,
\begin{equation}
\nu(M_{14}=12,z)=\frac{2.8}{D(z)}.
\end{equation}
Finally, one obtains for $D(1.39)=0.53$ that $\nu=5.3$. We find $\nu=5.0$ using our numerical code.

We are seeking to find the value of $\fnl$ necessary to increase the probability of this mass bin to a value indistinguishable from one. To be conservative we will stop at $P=0.95$. For a Poissonian distribution to have a probability of $0.95$ that there is at least one event, the expected number of events will need to exceed 3. Thus, we need an $\fnl$ that will increase the expected number of clusters belonging to the mass bin by a factor of 3/0.002=1500.

If we put this into eq.~(\ref{eq:numvj}), we find
\begin{equation}
 \ln(1500)=\frac{5.3^3}{6} 3\times10^{-4} \fnl,
\end{equation}
which gives $\fnl\simeq 1000$, a factor of two larger than that found in ref.~\cite{toobigtooearly}. Although finding a value of $\fnl$ much closer to this, ref.~\cite{CayonGordonSilk} are not safe. This is due to them having a much smaller survey window of 11 sq.~degrees. This will result in a much smaller probability for the cluster to exist in the Gaussian case, hence increasing the value of $\fnl$ required to make this mass bin probable.

\subsection{Discussion of Gaussian mass functions}\label{sec:Gaussmass}

Following \cite{toobigtooearly,CayonGordonSilk}, we have used above the Gaussian mass function from Jenkins et al.~\cite{Jenkins}.  The Jenkins
et al.~result used in ref.~\cite{toobigtooearly} is a fit to N-body simulations, which corresponds to
$\Lambda$CDM and the spherical overdensity group finder, and is given by
\begin{equation}
n_\mathrm{G}(M,z) = \frac{\bar{\rho}}{M} f \left( -\frac{\mathrm{d}\ln\sigma_M}{\mathrm{d}\ln M}\right)
\end{equation}
with
\begin{equation}\label{eq:Jenkins}
f = 0.301 \exp \left[ -|\ln \sigma_M^{-1}(z) +0.64|^{3.82} \right] \; .
\end{equation}

In ref.~\cite{Jenkins} this (and other) mass functions were tested and found to match well to simulations up to a value of $\ln\sigma^{-1}<1$. At redshift $z=0$ this corresponds to a mass of $5\times 10^{15} M_\odot/h$, which is well above the mass of any clusters we are considering. However, at redshift $z=1.4$, a mass of $10^{15}M_\odot$ would correspond to $\ln\sigma^{-1}=1.2$. As one probes deeper into redshift space, this mass function needs to be used at values of $\ln\sigma^{-1}$ further and further beyond those for which it is tested. 

This does not \emph{necessarily} mean the mass function \emph{will} be unusable outside of this range. The clusters themselves are not far outside of the tested range of these mass functions so it could be assumed that any errors introduced will be small if one applies eq.~(\ref{eq:defExp}) up to these masses and redshifts, but no further. However the redshift range probed by the cluster surveys is assumed in both refs.~\cite{toobigtooearly,CayonGordonSilk} to extend to $z=2.2$. Moreover, when considering the upper mass bin described in section \ref{sec:fnlest}, in principle one is required to integrate the mass to infinity. In practice this means integrating to some high mass where the \emph{full} cluster mass function is assumed to be negligible. For this assumption to hold, it is imperative that at these redshifts and masses the Gaussian mass function decreases fast enough with increasing $\sigma^{-1}$, since the non-Gaussian contribution is growing exponentially, with $\mathcal{R}\sim\exp(10^{-4}\fnl\sigma^{-3})$.

Once either $\fnl$ or $\sigma^{-1}$ grows large enough eq.~(\ref{eq:Jenkins}) will not decrease fast enough. In figure \ref{fig:jenkdown} we have plotted the full mass function at redshift $z=2.2$ for three values of $\fnl$. The mass function is weighted by the volume element $dV/dz$ to give the values on the y-axis more intuitive meaning. Each curve is effectively a plot of the integrand of eq.~(\ref{eq:defExp}) against mass for various $\fnl$ values. It is clear that the formalism breaks down at a particular mass for a given redshift and $\fnl$. The result of this breakdown is a very rapid turn-up in the full mass function at this mass. It is this effect and not a physical increase in the probability of clusters themselves existing that causes the sudden turn-up in the figures of refs.~\cite{toobigtooearly,CayonGordonSilk}.

In ref.~\cite{toobigtooearly}, the upper limit of the integral in eq.~(\ref{eq:defExp}) was set to $1.5\times10^{16} M_\odot/h$.\footnote{Private correspondence.} They also see the sudden turn-up in probability at $\fnl\simeq500$. On figure \ref{fig:jenkdown} we have drawn a line upwards at $M\simeq 1.5\times 10^{16} M_\odot/h$. It is very clear from where it crosses the $\fnl=500$ curve that $\fnl=500$ is the first $\fnl$ value where this breakdown in the mass function will be seen at this upper limit. The different value of $\fnl$ in ref.~\cite{CayonGordonSilk} for seeing this turn-up will be a result of a different upper limit used in eq.~(\ref{eq:defExp}). For context, a mass of $1.5\times10^{16}M_\odot/h$ at a redshift of $2.2$ corresponds to a value of $\sigma^{-1}=10.8$, which is well outside the range quoted as valid for the Jenkins mass function, eq.~(\ref{eq:Jenkins}).

\EPSFIGURE[t]{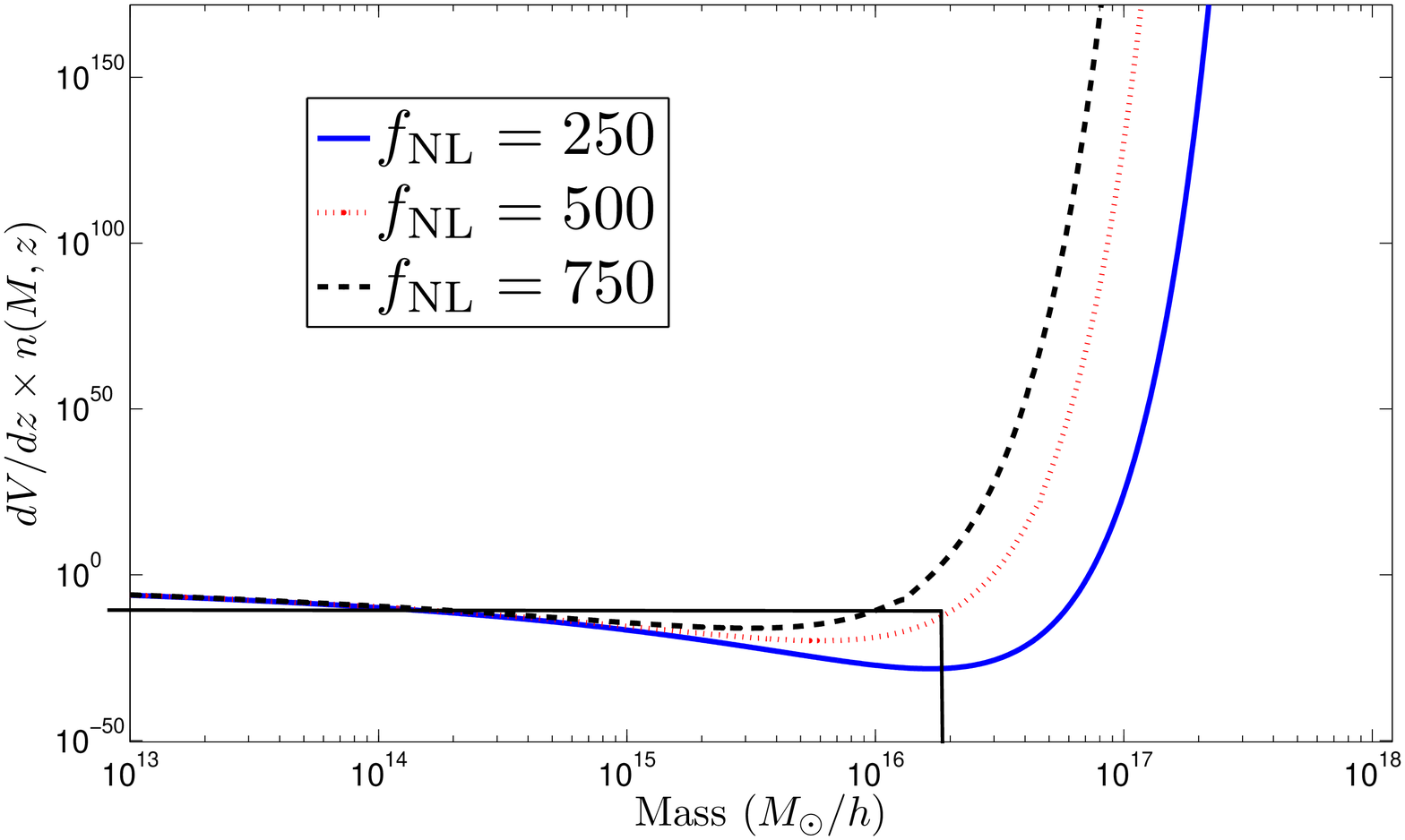,width=12cm}{\label{fig:jenkdown}Evidence of the breakdown of the non-Gaussian mass function at large masses when using the ratio method and the Jenkins et al. Gaussian mass function. Plot is of $dV/dz \times n(M,z)$ versus mass, at redshift $z=2.2$ and for some characteristic values of $\fnl$.}

To avoid this problem it is necessary to either use a Gaussian mass function that scales correctly at these high masses and redshifts, to use a full non-Gaussian mass function instead of the ratio method of eq.~(\ref{eq:ratio}) or to cut off the integral before this breakdown occurs. Unfortunately no N-body fits have been tested this far into the tail of the distribution in both mass and redshift. In the absence of a tested function, we can ask what asymptotic behaviour we expect from theory. Thankfully, although differing in exact functional form, all theoretical models reproduce the same behaviour for the Gaussian mass function at large masses and redshifts, with $f\sim \exp(-c/\sigma^2)$ in equation (\ref{eq:Jenkins}). The form of eq.~(\ref{eq:Jenkins}) makes it very difficult to judge how it will behave asymptotically as compared with $\exp(-c/\sigma^2)$. Therefore, we will use the formula from Tinker et al.~\cite{Tinkeretal}\footnote{This very simple formula matches the more complicated red-shift dependent formula presented in the update \cite{Tinker:2010my} to within 5\%. We use this formula due to its clear and desired asymptotic behaviour for small $\sigma$.}, given by 
\begin{equation}\label{eq:Tinker}
f(\sigma)=A\left[ \left(\frac{\sigma}{b}\right)^{-a}+1\right]e^{-c/\sigma^2} \; ,
\end{equation} 
where for an spherical overdensity $\Delta=200$ they find $A=0.186$, $a=1.47$, $b=2.57$, and $c\simeq1.19$. This numerical fit also matches the most sophisticated theoretical models over a range of $\sigma^{-1} = 0.5 \rightarrow 3$, to within the errors introduced in the theoretical models \cite{MaggioreRiotto2}.

\DOUBLEFIGURE[t]{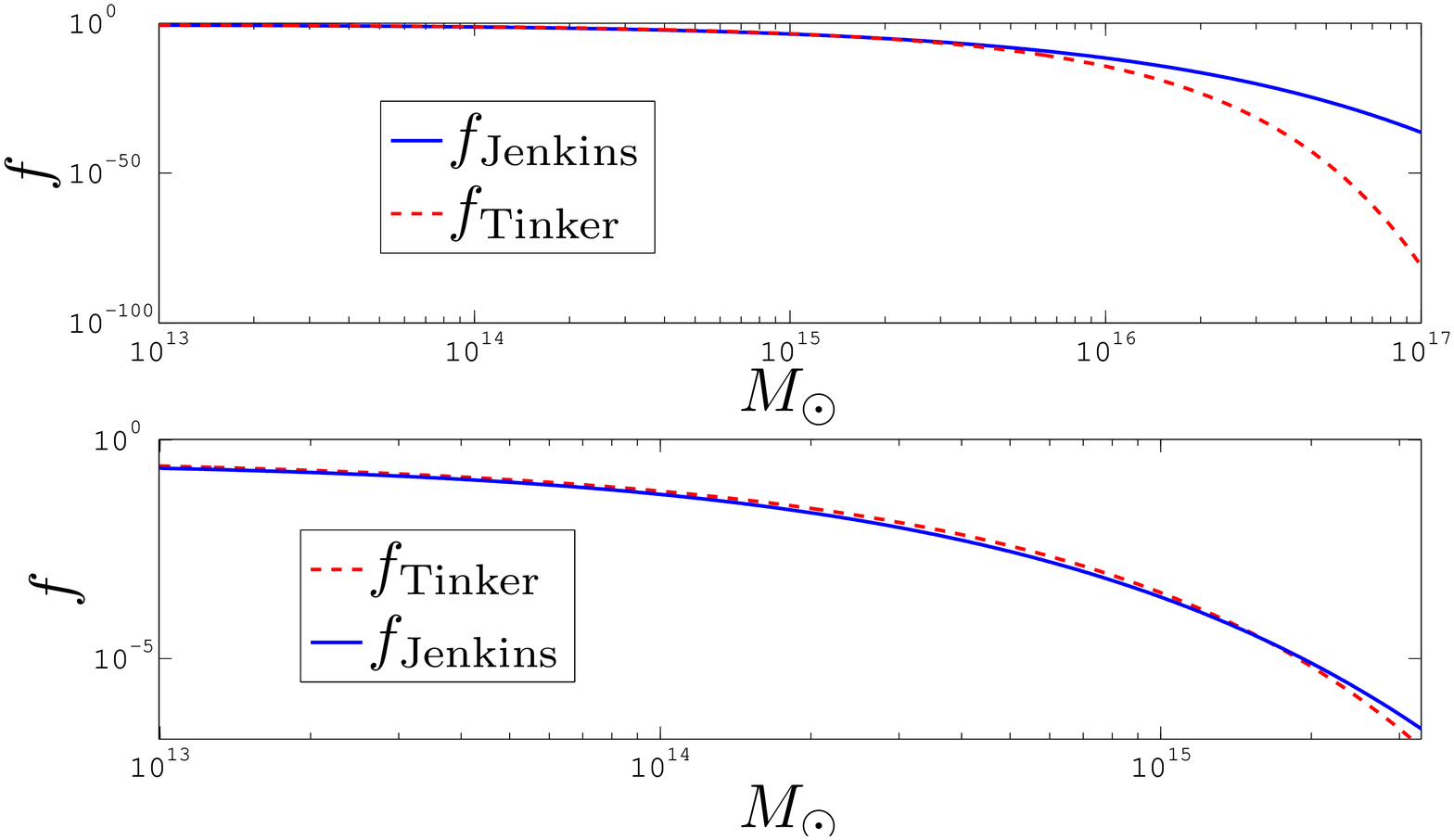,width=8cm}{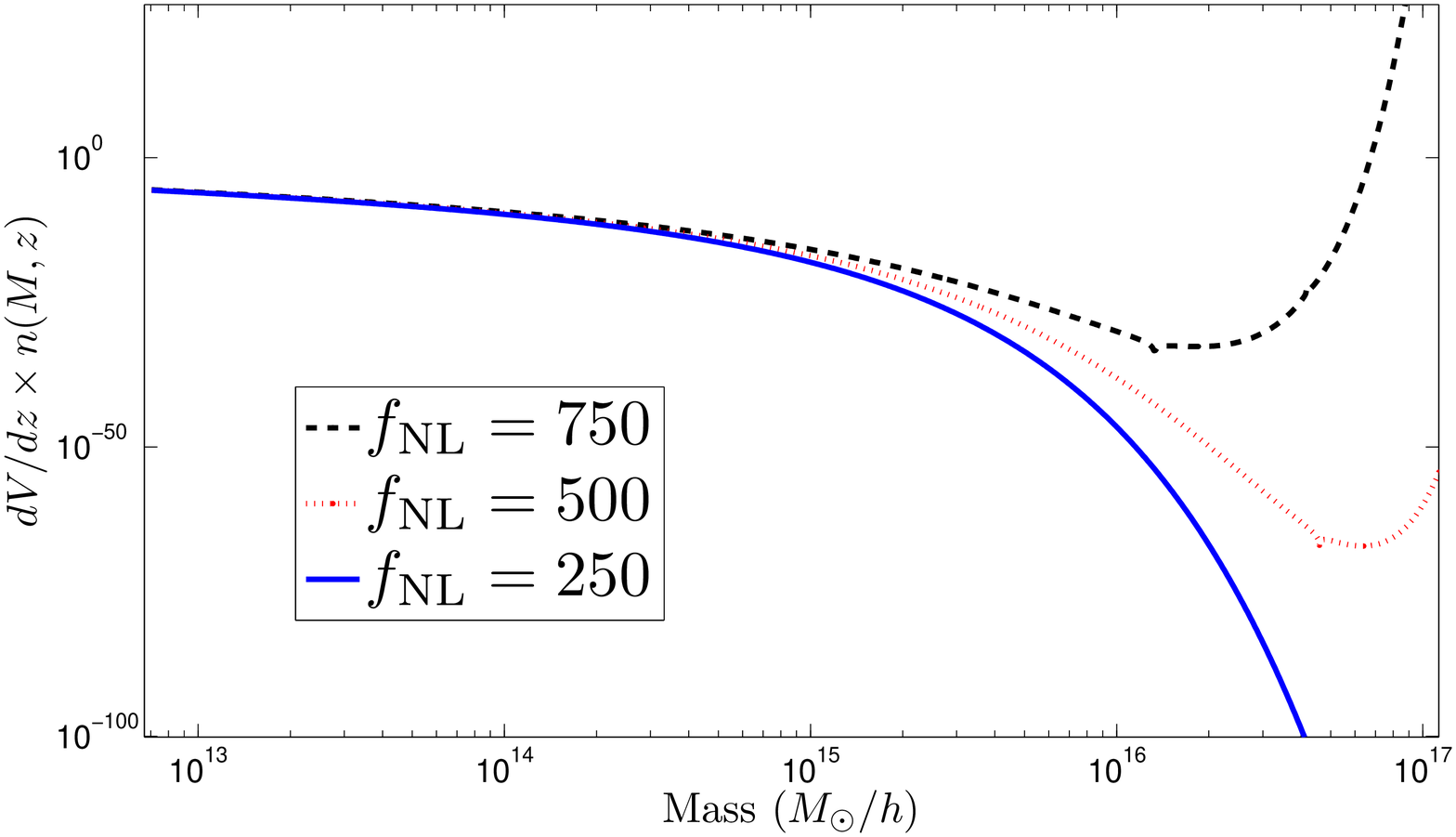,width=8cm}{\label{fig:massfunctions} Comparison of the Gaussian mass functions from Jenkins et al.~\cite{Jenkins} and Tinker et al.~\cite{Tinkeretal} plotted against mass, for $z=1$. The lower plot is simply a zoomed in version of the upper plot.}{\label{fig:tinkdown} Evidence of the breakdown of the non-Gaussian mass function at \emph{very} large masses using Tinker. Plot is of $dV/dz \times n(M,z)$ versus mass, at redshift $z=2.2$ and for the same $\fnl$ values as fig.~\protect\ref{fig:jenkdown}.}

Figure \ref{fig:massfunctions} is a comparison of both mass functions. At smaller masses they match closely but diverge from each other at larger masses. This indicates that the Jenkins et al.~mass function cannot be behaving asymptotically as $f\sim\exp(-c/\sigma^2)$. This is the origin of the problematic behaviour seen in figure \ref{fig:jenkdown}. In figure \ref{fig:tinkdown} we plot the analogue of figure \ref{fig:jenkdown} using the Tinker et al.~mass function. Two things are immediately apparent. Firstly, close to the relevant cluster masses it is much better behaved than the full mass function derived using the Jenkins et al. Gaussian--- this  is exactly what we expected from figure \ref{fig:massfunctions}. Secondly, it also breaks down if probed to large enough masses. This indicates that even a mass function that asymptotically matches the best theoretical models at both the Gaussian and non-Gaussian limits, breaks down at large enough masses. This too, however, is not unexpected and will occur even for the full theoretical non-Gaussian mass functions that do not use the ratio method of eq.~(\ref{eq:ratio}). This is discussed briefly in \cite{D'AmicoMussoNorenaParanjape}. This breakdown occurs at the point that $\nu (\sigma S_3) \simeq 1$ which is the point where the non-Gaussian contribution to the density perturbations becomes comparable in size to the Gaussian contribution. For the least probable clusters so far detected, $\fnl\simeq1000$ is sufficiently large for the breakdown to occur. To set an upper bound on $\fnl$ in the future it will be necessary to extend the non-Gaussian mass function to $\nu (\sigma S_3) \gtrsim 1$.

With a clear understanding of the limitations of the two mass functions, it is pertinent now to perform the analysis of section \ref{sec:fnlest} numerically. The results are presented in figures \ref{fig:fnljenkins15}, \ref{fig:fNLjenkins45} and \ref{fig:fnltinker} for three of the clusters in table \ref{ourtable}. Figure \ref{fig:fnltinker} is the correct plot using the Tinker et al.~mass function. By comparing figures \ref{fig:fnljenkins15} and \ref{fig:fNLjenkins45} it is clear that, with the Jenkins et al.~mass function, it is possible to cause the sudden, unphysical, jump in the probability of the upper mass bin to occur at any arbitrary value of $\fnl$. This would be done by tuning the cutoff of the integral in eq.~(\ref{eq:defExp}) to cause the unphysical turn-up of figure \ref{fig:jenkdown} to creep into the integrated range at precisely the chosen value of $\fnl$. More positively however, it is also clear from examining all three figures, that until this unphysical jump does enter the integral, the results are consistent. This gives us confidence that the results presented in refs.~\cite{toobigtooearly,CayonGordonSilk} will be correct whenever they correspond to $\fnl$ beneath this critical of $\fnl$ for their respective choice of $M_\mathrm{max}$. For ref.~\cite{toobigtooearly} this will be $\fnl\lesssim500$, for ref.~\cite{CayonGordonSilk} this will be $\fnl\lesssim700$.

\EPSFIGURE[t]{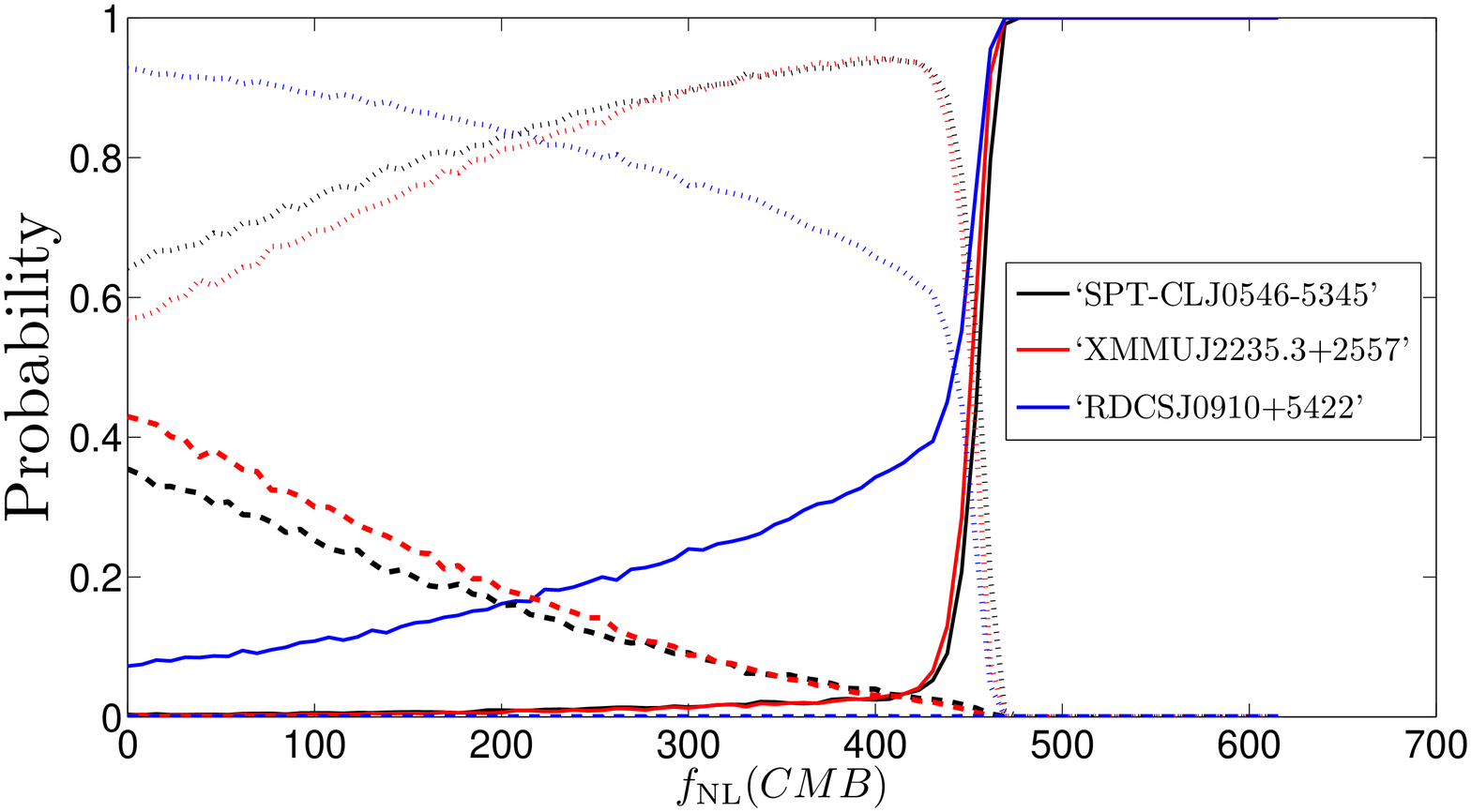,width=12cm}{\label{fig:fnljenkins15} The probability that each of the three labelled clusters could exist and be the ``most massive'' cluster in the survey using the Jenkins et al.~mass function. Following the convention of \cite{toobigtooearly}, the solid line depicts the probability that the ``most massive'' cluster in the survey is more massive than the labelled cluster, the dotted line depicts the probability that the ``most massive'' cluster's mass falls within the 1-$\sigma$ error range of the labelled cluster and the dashed line depicts the probability that the ``most massive'' cluster is  less massive than the labelled cluster. For this curve we set $M_\mathrm{max}=1.5\times 10^{16} M_\odot/h$ in equation (\protect\ref{eq:defExp}).}

\EPSFIGURE[t]{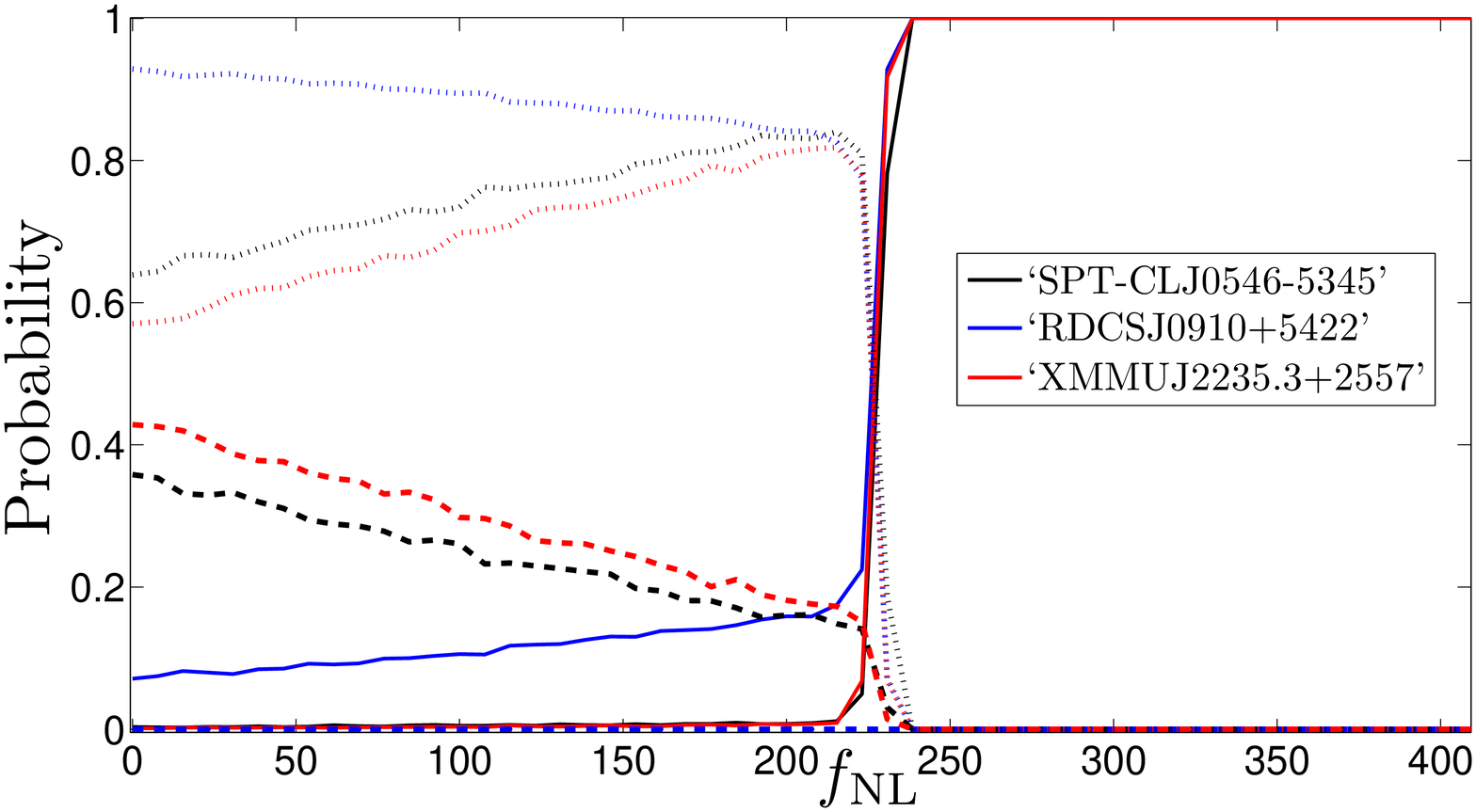,width=12cm}{\label{fig:fNLjenkins45} This is an identical figure to figure \protect\ref{fig:fnljenkins15}, except we set $M_\mathrm{max}=4.5\times 10^{16} M_\odot/h$ in eq.~(\protect\ref{eq:defExp}). This is clear evidence of the cutoff dependence of the results due to the effect seen in figure \protect\ref{fig:jenkdown}.}
%

\EPSFIGURE[t]{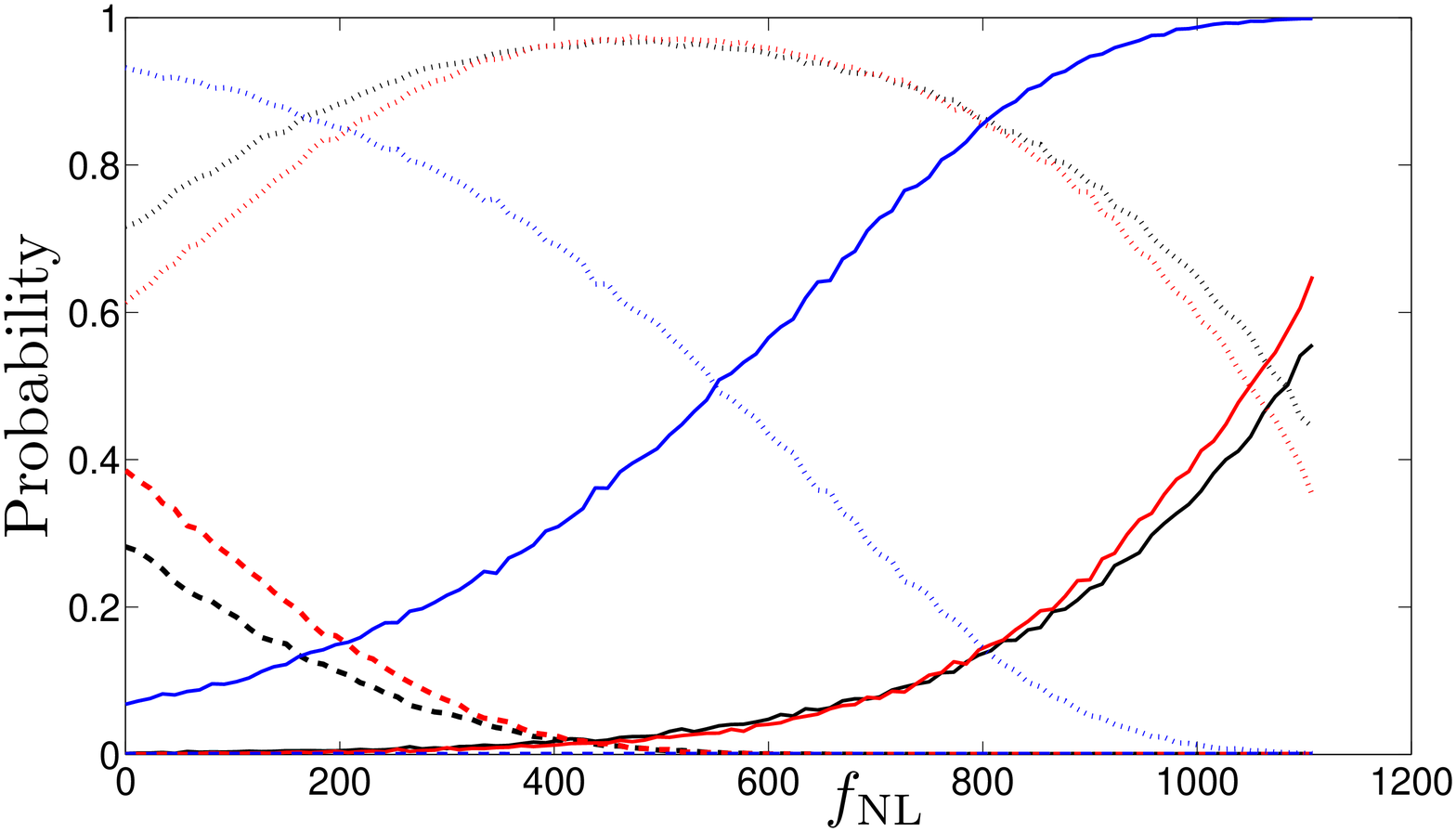,width=12cm}{\label{fig:fnltinker} This figure is identical to figures \protect\ref{fig:fnljenkins15} and \protect\ref{fig:fNLjenkins45} except it has been calculated using the Tinker et al.~mass function. These results are insensitive to the choice of $M_\mathrm{max}$.}

\EPSFIGURE[t]{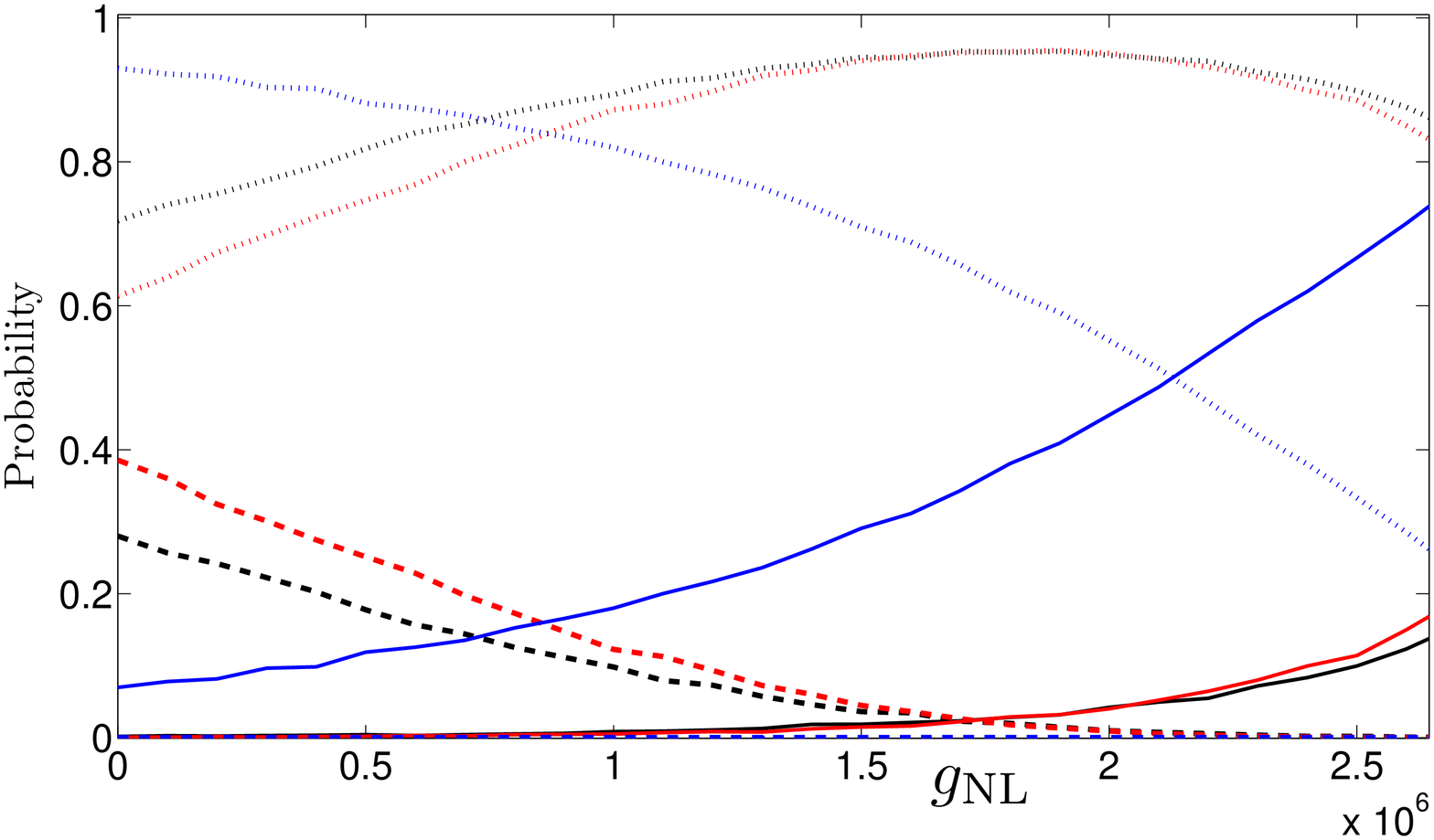,width=12cm}{\label{fig:gnltinker} This figure is an exact analogue of figure \protect\ref{fig:fnltinker}, but for $\gnl$. The solid curves depict the probability that the `most massive cluster' is greater than the labelled cluster, the dotted curves depict the probability that the `most massive' cluster fits within the 1-$\sigma$ error range of the labelled cluster and the dashed line depicts the probability that the `most massive' cluster is less massive than the labelled cluster. The appropriate labels are present on figure \protect\ref{fig:fnljenkins15}.}
%

\subsection[Robust numerical results for $\fnl$]{Robust numerical results for \boldmath$\fnl$}
\label{sec:numfnl}

We now seek to use the entire ensemble of 15 clusters to constrain $\fnl$. For results presented in this section we always calculate the skewness, $S_3$, fully numerically. We follow the method of ref.~\cite{toobigtooearly} very closely including using their \emph{conservative} estimates for $f_\mathrm{sky}$. For $\fnl<500$ we find similar results. We denote ${\bf M}_i$ as the random variable giving the mass of cluster $i$ and we denote $M_i$, $\sigma^+_i$ and $\sigma_i^-$ as the mass, upper error and lower error respectively of cluster $i$ quoted in table \ref{ourtable}. We take $\ln{\bf M}_i$ to be normally distributed with a mean of
\begin{equation}
\mu_i=\ln M_i.
\end{equation}
Also, we take the standard deviation, $\sigma_i$, of $\ln{\bf M}_i$ to be
\begin{equation}
\sigma_i=\frac{1}{2}\ln\left( \frac{M_i+\sigma_i^+}{M_i-\sigma_i^-}\right).
\end{equation}
For each of the 15 clusters, we then sample ${\bf M}_i$ $10^4$ times from this distribution. For each of the $10^4$ mass samples we calculate the total number of clusters expected in the survey window at equal or higher mass and redshift. We then Poisson sample over each one of these expectation values. The total number of these samples that return a value greater than zero allows us to form a probability, $P_i$, that the given cluster can exist, marginalised over its assumed log Gaussian distribution. We then multiply the individual cluster probabilities together to form the probability, $P(\fnl)=\prod P_i$, that the full ensemble could exist in the survey window. Finally, we repeat the above analysis for increasing values of $\fnl$ and record the value of $\fnl$ when $P(\fnl)=0.05$.

\DOUBLEFIGURE[t]{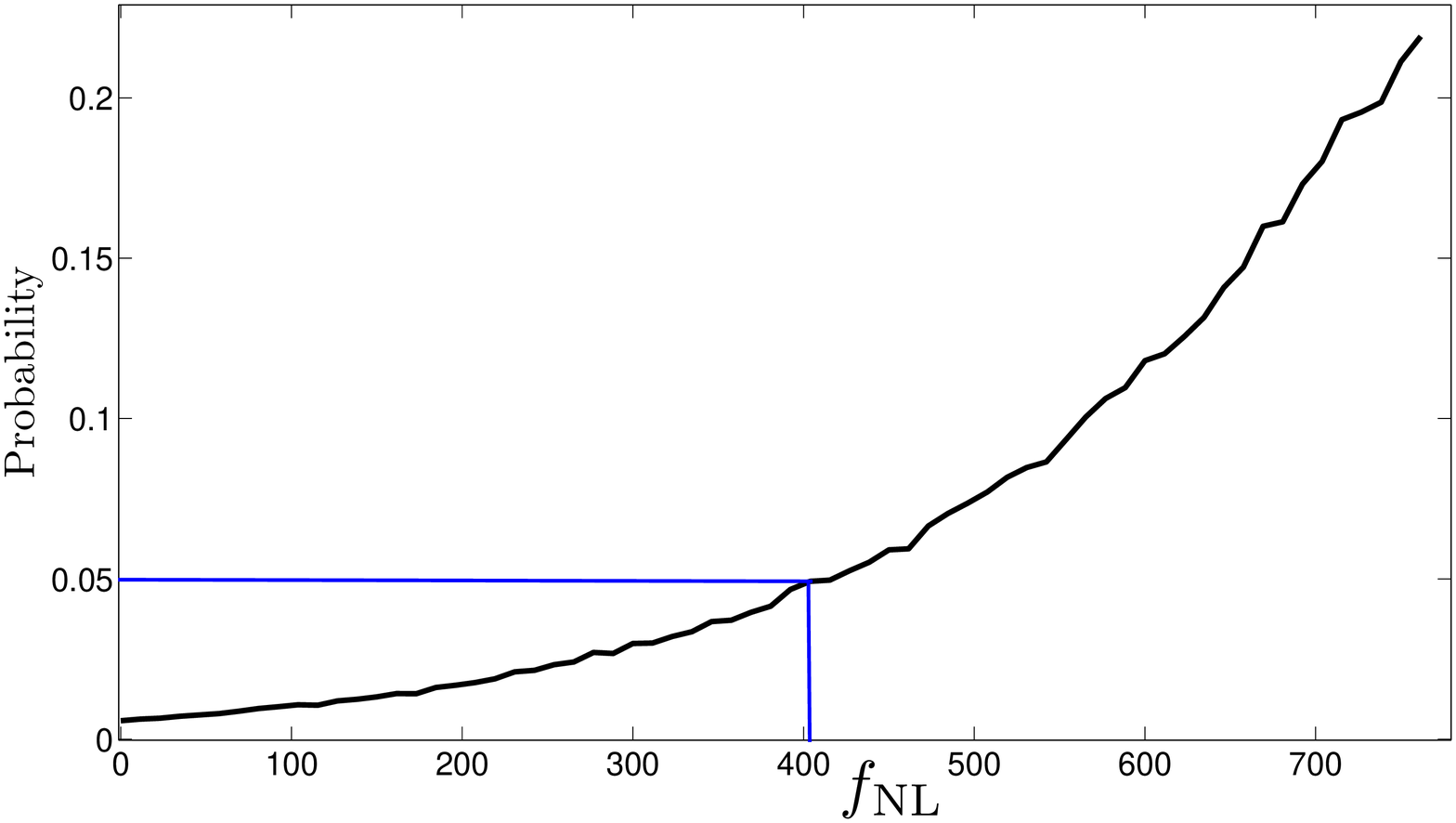,width=8cm}{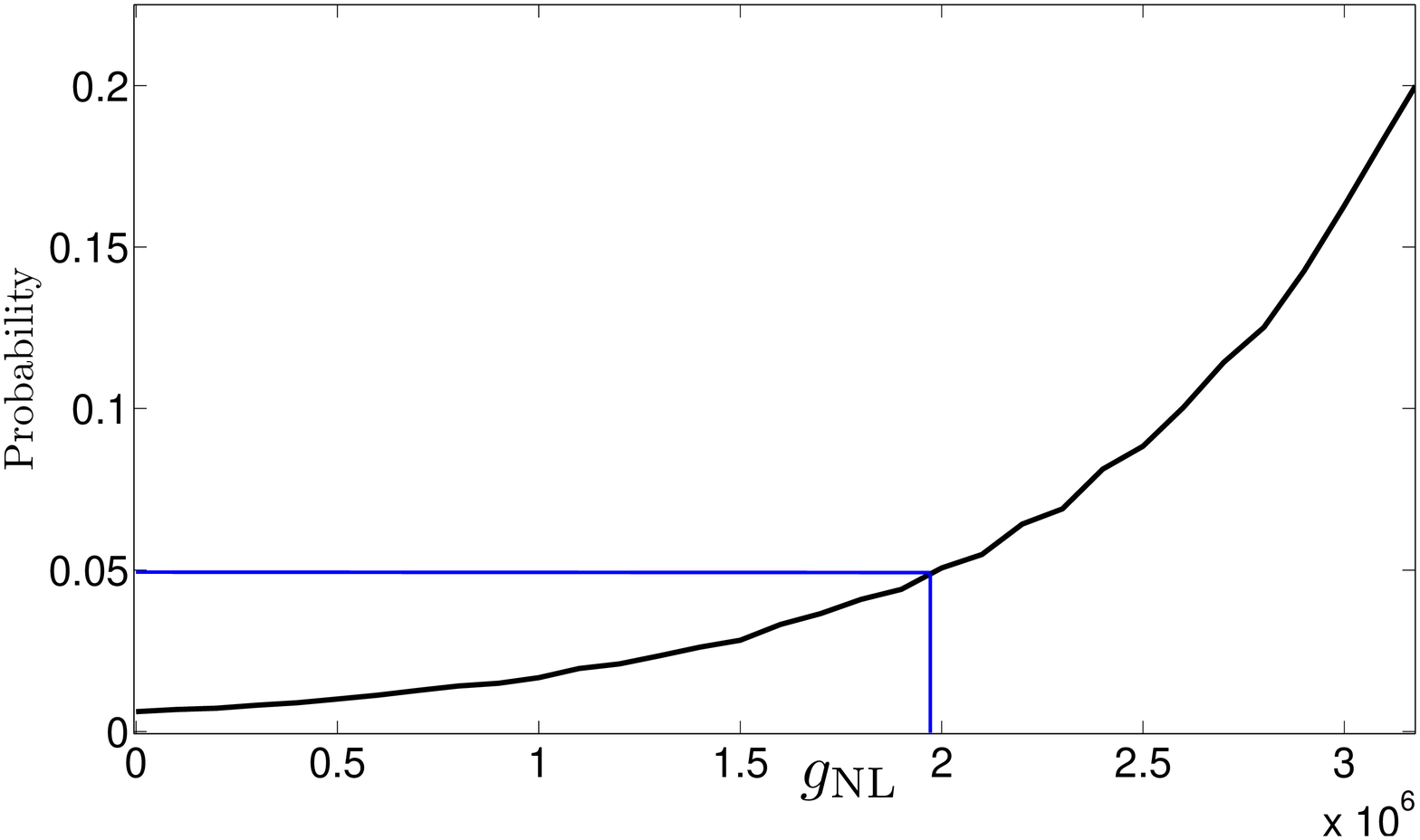,width=8cm}{\label{fig:fnlfull} The probability that the ensemble of clusters in table \protect\ref{ourtable} could exist as a function of $\fnl$.}{\label{fig:gnlfull} The probability that the ensemble of clusters in table \protect\ref{ourtable} could exist as a function of $\gnl$, with $\fnl\lesssim 50$.}

In figure \ref{fig:fnlfull} we have plotted $P(\fnl)$. If we compare this to figure 4 of ref.~\cite{toobigtooearly} we see that the two results match well until $\fnl\simeq500$. From the data used to produce this figure we find a lower bound on $\fnl|_{P=0.05}$ with WMAP 5 year parameters of $\fnl|_{P=0.05}>412$ required to make the existence of this ensemble of clusters consistent with $\Lambda$CDM. ref.~\cite{toobigtooearly} quote the corresponding value to be $\fnl|_{P=0.05}>476$. This is close, but perhaps more different than we would initially expect. We suggest the differences will come from different definitions of $\mu$ and $\sigma$; although it could also potentially come from our use of ref.~\cite{Bardeen:1985tr}'s fitting formula for the transfer function, compared to ref.~\cite{toobigtooearly} who numerically integrate theirs using the icosmo package \cite{Refregier:2008fn}. Ordinarily one would not expect $>10\%$ errors; however a systematic $<10\%$ error in all 15 clusters could accumulate into a discrepancy of this size in $P(\fnl)$. 

Our main motivation is to compare this lower bound for $\fnl$ to the one we will derive in section \ref{section:gnl} for $\gnl$. Therefore, for concision we will not marginalise this $\fnl$ constraint over the cosmological parameters. The results would be similar to that obtained by ref.~\cite{toobigtooearly}, except for regions where $\fnl>500$, as discussed earlier. 

The analysis in ref.~\cite{CayonGordonSilk} uses a slightly different method to calculate their final constraints on $\fnl$, but gain a similar value, quoting $\fnl=449\pm286$. Their quoted numbers, however, are more dependent on the spurious behaviour caused by the breakdown of the mass functions, so their results change more significantly when this is removed. Nevertheless, the effect of removing it would be to make larger $\fnl$ more probable. This would increase the central value of $\fnl$ that they quote, making their conclusions about non-zero and scale dependent $\fnl$ even stronger. Both a marginalisation over cosmology and a re-analysis similar to ref.~\cite{CayonGordonSilk} would be interesting future calculations.

\section{Analytic estimate and numerical results for \boldmath$\gnl$}
\label{section:gnl}

We now repeat the calculations of the previous section, but for $\gnl$, assuming $\fnl$ is small enough to be negligible in comparison (in practice this means $\fnl\lesssim50$). A positive $\fnl$ would mildly reduce our quoted lower bound on $\gnl$ and a negative $\fnl$ would mildly increase it.

\subsection[Estimate of $\gnl$ required to give similar effect as $\fnl$]{Estimate of \boldmath$\gnl$ required to give similar effect as $\fnl$}

To estimate the effect of $\gnl$ we use the $\gnl$ dependent part of eq.~(\ref{eq:MVJfnl}), which in terms of $\nu$, is to a good approximation
\begin{equation}
 \mathcal{R}\simeq\exp\left(\frac{\nu^4}{24} (\sigma^2 S_4)\right).
\end{equation}
This equation appears in the calculation for $\gnl$ in exactly the same manner as eq.~(\ref{eq:numvj}) does for $\fnl$.  Therefore, in order for a given $\gnl$ to cause a similar sized effect to a given $\fnl$ it is necessary that
\begin{equation}
 \frac{\nu^4}{24} (\sigma^2 S_4)\simeq\frac{\nu^3}{6} (\sigma S_3)
\end{equation}
or,
\begin{equation}
 \gnl=\frac{2\times10^4}{\nu} \fnl.
\end{equation}
For our least probable cluster, with $\nu\simeq4.5$, this gives
$\gnl\simeq 4.4\times 10^3 \fnl$.  The least probable
clusters dominate the departures from
$\Lambda$CDM, therefore to a reasonable approximation we can use this
result to relate our earlier constraints on $\fnl$ to $\gnl$. We found
earlier that $\fnl\simeq 410$ was sufficient to give the ensemble of
clusters a probability of 0.05 of existing. We therefore expect for
$\gnl\simeq 1.8\times 10^6$ the same to be true. As we will see, this
is indeed the case.

\subsection[Numerical $\gnl$]{Numerical \boldmath$\gnl$}\label{sec:numgnl}

Here we present the numerical results for $\gnl$ using precisely the same methodology that was used for $\fnl$ in section \ref{sec:numfnl}. In \cite{DesjacquesSeljak2009} it was found that in the mass range $10^{13}\lesssim M \lesssim 5\times10^{15}\,\, M_\odot/h$, $\sigma^2 S_4$ varies in the narrow range $\sim 4-6 \times 10^{-8} \gnl$ (note that this particular combination, $\sigma^2 S_4$, is independent of redshift). This matches closely to our analytical estimate, $\sigma^2 S_4 \simeq 5.8\times10^{-8} \gnl$. Therefore, we choose to trust the results of ref.~\cite{DesjacquesSeljak2009} and take $\sigma^2 S_4 = 5\times10^{-8} \gnl$ to be constant over all redshifts and the mass range $10^{13}\lesssim M \lesssim 5\times10^{15}\,\, M_\odot/h$. This range comfortably encompasses all of the clusters in table \ref{ourtable}. We expect this approximate method of defining $\gnl$ should introduce errors in $\gnl$ of $\lesssim20\%$.

Due to the more rapid scaling of $\mathcal{R}$ with $\nu$ for $S_4$ as compared to $S_3$, we noticed that at $z=2.2$ and with $M_\mathrm{max}\gtrsim10^{16}M_\odot$, the full mass function sometimes broke down within the range of our integral. This occurs even with the Gaussian mass function from Tinker et al.~\cite{Tinkeretal}. This is due to $\nu^2 (\sigma^2 S_4)$ becoming of order one, rather than due to a breakdown in the ratio method. To avoid this affecting our results, we imposed a cutoff in our integral at $\nu=7.2$. This is comfortably larger than values near any of the cluster masses and redshifts. It is also small enough that the mass function does not break down for any interesting values of $\gnl$. We tested the robustness of this approximation by varying the arbitrary cutoff. Within the ranges quoted and plotted, our results did not change. The changes that did occur were for very large $\gnl$ precisely where we would expect the mass function to break down near $\nu=7.2$.

In figure \ref{fig:gnltinker} we show the plot of $\gnl$ that is equivalent to figure \ref{fig:fnltinker}. We see very similar behaviour to the plot of $\fnl$. We would also see similar behaviour to figures \ref{fig:fnljenkins15} and \ref{fig:fNLjenkins45} if we were to use the Gaussian mass function from Jenkins et al. In figure \ref{fig:gnlfull} we show the plot of $\gnl$ that is equivalent to figure \ref{fig:fnlfull}. From this figure we see that our estimate of $\gnl\simeq 1.8\times 10^{6}$ is very accurate. When we extract the precise value at which $P(\gnl)=0.05$, we find $\gnl|_{P=0.05}>2.0\times10^6$. If we compare this lower constraint to the upper constraints listed in the introduction, we note a small degree of tension. However the amount of tension is much smaller than in the $\fnl$ case.

\EPSFIGURE[t]{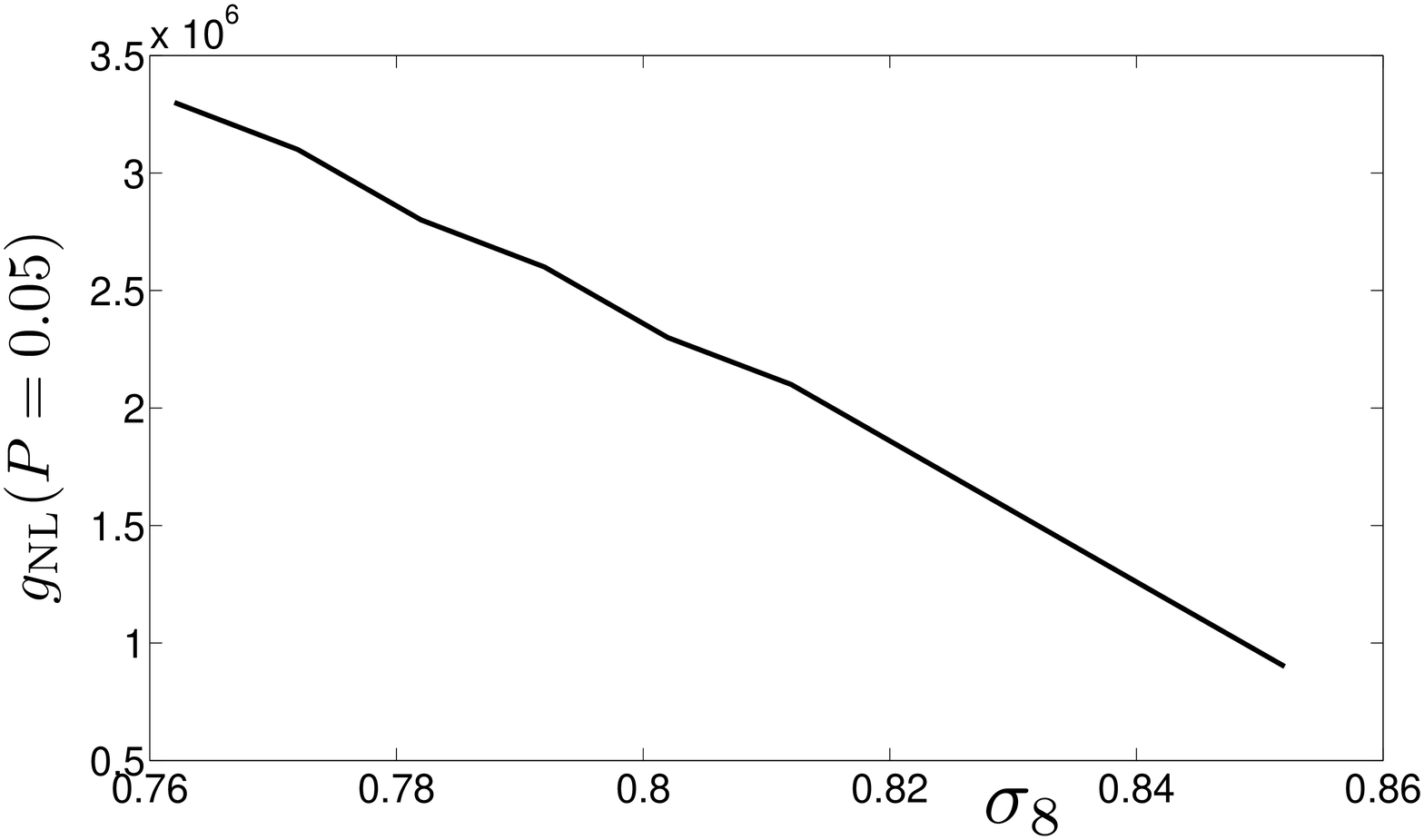,width=12cm}{\label{fig:gnlsig8} The effect of varying $\sigma_8$ on our $\gnl$ estimate (i.e. the value of $\gnl$ for which the ensemble of clusters in table \ref{ourtable} exists with probability $P=0.05$).}

Finally, in figure \ref{fig:gnlsig8}, we give a plot of $\gnl|_{P=0.05}$ against $\sigma_8$. To calculate the effect of changes in $\sigma_8$ on $S_4$ we multiplied $\sigma^2 S_4$ by the ratio $\sigma_8^2/0.81^2$ (note: $\sigma_8=0.81$ was the value for which ref.~\cite{DesjacquesSeljak2009} calculated their value for $\sigma^2 S_4$). This takes into account the effect of $\sigma_8$ on $S_4$ exactly. This is because $\sigma_8$ only defines the normalisation of $\mathcal{P}$ and does not enter into any scale dependent quantities in the integral defining $S_4$. The important result of this figure is that for large enough $\sigma_8$, but still within observational constraints, $\gnl|_{P=0.05}$ itself can be brought within current observational constraints. This would suggest that we can fully explain the existence of these massive, high-redshift clusters without appealing to scale dependent non-Gaussianity. This is not the case for $\fnl$. 

This `suggestion' might in fact be true, but we caution against taking it too seriously. The constraints on $\gnl$ quoted in the introduction were obtained for a value of $\sigma_8\simeq0.81$. If we increase $\sigma_8$, then the degeneracy we see here that allows a smaller $\gnl$ will force a smaller $\gnl$ in those constraints as well. The open question is which degeneracy is more sensitive. This would be an interesting topic to pursue in the future.

\section{Conclusions}
\label{section:conclude}
We have found and remedied a common problem in the literature \cite{toobigtooearly,CayonGordonSilk} that occurs when estimating $\fnl$ from massive high redshift clusters. We found that the value of $\fnl$ necessary to make the existence of these massive clusters certain (i.e. to give 100\% confidence that they would be detected in the given surveys) is $\fnl \gtrsim 1000$, which is significantly larger than what has been claimed (e.g.~$\fnl > 550$ in \cite{toobigtooearly}). We have also demonstrated that this discrepancy is explained by the fact that \cite{toobigtooearly} and \cite{CayonGordonSilk} use a non-Gaussian mass function, derived from a Gaussian mass function that is not valid in the whole range of $\ln\sigma^{-1}$ needed for the computation. Most importantly the Gaussian mass function does not scale correctly outside of this range. We have rectified this oversight by using a Gaussian mass function by Tinker et al.~\cite{Tinkeretal} that has the proper asymptotic behaviour. We also found that below a certain critical $\fnl$ value, the results in both refs.~\cite{toobigtooearly,CayonGordonSilk} can be trusted with confidence. However, this value differs for each references and depends on an arbitrary cutoff of the mass integral in equation (\ref{eq:defExp}). Our 95\% confidence lower bound of $\fnl\gtrsim410$ agrees reasonably well with reference \cite{toobigtooearly} because it occurs below this critical $\fnl$ value for their cutoff.

We then considered the case where the non-Gaussian part of the
primordial spectrum is  dominated by $\gnl$. 
Such a situation can easily arise e.g.~in curvaton models \cite{Curvaton1, Curvaton2, Curvaton3}.
We estimated $\gnl > 2.0 \times 10^6$. We thus demonstrated, that within current observational limits, $\gnl$ appears to have
more potential to explain the observed excess of high-redshift massive
clusters.

Non-Gaussianity is not the only potential explanation for the tension caused by the existence of these high redshift clusters. A systematic error that consistently over-estimated the masses would have a similar effect. In \cite{toobigtooearly} this possibility was considered. To minimise the effects of systematic errors, the mass measurements used were always taken to be those whose quoted errors were consistent with the smallest mass value. We use the same mass estimates. For many clusters, this involved comparing mass estimates from SZ effect measurements, X-ray measurements and weak lensing measurements. Any systematic errors would need to be present in all three measurement methods. It was also found in \cite{toobigtooearly} that all the masses would need to be systematically over-estimated by 1.5 $\sigma$ in \emph{each} measurement technique to make this ensemble of clusters fully consistent with $\fnl=0$.

A different expansion history is another potential explanation for the existence of these clusters \cite{Alam:2010tt,Bhattacharya:2010wy,Xia:2009ys}. A modified equation of state for dark energy that resulted in an earlier onset of the accelerated expansion would suppress structure growth at smaller redshifts/later times. This would have the effect of causing the linear growth function $D(z)$ to drop more slowly from low redshifts to high redshifts. The net result is more structure and thus more, large mass, clusters at high redshifts than what would be expected in $\Lambda$CDM.

The estimates of $\fnl$ quoted here and in \cite{toobigtooearly} and
\cite{CayonGordonSilk} are far outside the observational limits set by WMAP. In
\cite{toobigtooearly} and \cite{CayonGordonSilk} they suggested remedying this problem by
introducing running of $\fnl$. While this would explain the apparent
discrepancy of the magnitude of $\fnl$ over different scales, it
is also likely to introduce problems with the actual spectral index of the
perturbations, since usually if $\fnl$ acquires running, so does the
magnitude of the perturbations. (For discussions on running non-linearity
parameters, see \cite{Scale1, Scale2, Scale3, Scale4, Scale5}.)

We have demonstrated that instead of introducing non-zero $n_{\fnl}$, the  abundance of the massive clusters can be explained by introducing $\gnl$ almost within the current observational bounds. We also like to draw attention to a potentially important observational result: there appears to be an overabundance of large voids (see \cite{Voids1,Voids2} and references therein), and while large $\gnl$ makes both heavy clusters and large voids more probable, a positive $\fnl$ actually makes the voids \emph{less} likely, increasing the tension with observations even more \cite{D'Amico:2010kh}. In general it seems that a large value of $\gnl$ is in much better agreement with all observations (CMB, clusters, halo bias, voids), than large values of $\fnl$. With better measurements and N-body simulations a slightly smaller value of $\gnl$ could perhaps accommodate the abundance of the heavy clusters. This stresses the importance of extracting limits from the Planck microwave temperature map not only on $\fnl$ but also on $\gnl$.

\begin{acknowledgments}

We would like to thank Ben Hoyle for initial methodological help and subsequent discussions on the sources of the 
computational discrepancies. OT is supported by the Magnus Ehrnrooth foundation.
KE and SH are respectively supported by the Academy of Finland grants 218322 and 131454.

\end{acknowledgments}

\end{document}